\documentclass[amsmath,amssymb,pra,aps,showpacs,superscriptaddress,twocolumn]{revtex4-1}
\usepackage{amsmath,amsfonts,amssymb,amsthm,graphics,graphicx,epsfig,bbm}
\usepackage[colorlinks=true,citecolor=blue,linkcolor=blue,urlcolor=blue]{hyperref}
\usepackage[usenames]{color}
\usepackage{graphicx}
\usepackage{subfigure}
\usepackage{amsmath}
\usepackage{epsfig}
\usepackage{dcolumn}
\usepackage{bm}
\usepackage{color}
\usepackage{times}
\usepackage{epstopdf}
\usepackage{amssymb}
\usepackage{amstext}
\usepackage{latexsym}
\usepackage{hyperref}
\usepackage{amsfonts}
\usepackage{psfrag}
\usepackage{soul,xcolor}
\usepackage[normalem]{ulem}
\usepackage{dsfont}
\usepackage{txfonts}

\newcommand{\ket}[1]{\vert #1 \rangle}

\begin{document}
\setstcolor{red}
	
\title{Optimizing Counterdiabaticity by Variational Quantum Circuits} 
	
\author{Dan Sun}
\affiliation{International Center of Quantum Artificial Intelligence for Science and Technology (QuArtist) \\ and Department of Physics, Shanghai University, 200444 Shanghai, China\\}
	 
\author{Pranav Chandarana}
\affiliation{Department of Physical Chemistry, University of the Basque Country UPV/EHU, Apartado 644, 48080 Bilbao, Spain\\}
\affiliation{EHU Quantum Center, University of the Basque Country UPV/EHU, Barrio Sarriena, s/n, 48940 Leioa, Spain}
		
\author{Zi-Hua Xin}
\affiliation{International Center of Quantum Artificial Intelligence for Science and Technology (QuArtist) \\ and Department of Physics, Shanghai University, 200444 Shanghai, China\\}

\author{Xi Chen}
\email{chenxi1979cn@gmail.com}
\affiliation{Department of Physical Chemistry, University of the Basque Country UPV/EHU, Apartado 644, 48080 Bilbao, Spain\\}
\affiliation{EHU Quantum Center, University of the Basque Country UPV/EHU, Barrio Sarriena, s/n, 48940 Leioa, Spain}

\begin{abstract}
Utilizing counterdiabatic (CD) driving - aiming at suppression of diabatic transition - in digitized adiabatic evolution have garnered immense interest in quantum protocols and algorithms. However, improving the approximate CD terms with a nested commutator ansatz is a challenging task. In this work, we propose a technique of finding optimal coefficients of the CD terms using a variational quantum circuit. By classical optimizations routines, the parameters of this circuit are optimized to provide the coefficients corresponding to the CD terms. Then their improved performance is exemplified  in Greenberger-Horne-Zeilinger state preparation on nearest-neighbor Ising model. Finally,  we also show the advantage over  the usual quantum approximation optimization algorithm, in terms of fidelity with bounded time.

\end{abstract}

 	\maketitle

\section{Introduction}
Quantum computing has been of significant interest for a long time now due to the successes in the fields of machine learning~\cite{jacob2017,vedran2016}, quantum simulation~\cite{richard1999}, optimization \cite{harrigan2021,niko2018,edward2016}, and others. Especially, solving optimization problems with quantum computers as they are expected to outperform classical computers in the Noisy Intermediate-Scale Quantum (NISQ) era, see current review~\cite{reviewNISQ}. Among various methods~\cite{alberto2014,edward2014,jarrod2016,K.mitarai2018}, adiabatic quantum optimization (AQO)~\cite{Davide2019} is an effective method to tackle these problems. In AQO, the optimization problem is encoded into a  problem Hamiltonian such that the ground state corresponds to the solution of the problem. Then, a simple Hamiltonian, whose ground state is easy to prepare, is taken and adiabatically evolved to the problem Hamiltonian. The adiabatic theorem guarantees that the problem Hamiltonian will be in the ground state with a high success probability. The problem with this is that adiabatic processes are slow, and becomes unfeasible. With the digitized adiabatic evolution~\cite{Barends2016} , AQO can be reformulated into a gate model. However,  the long time evolution requires lots of gates, which spoils the desired results due to the errors in the noisy quantum devices.

To circumvent this, shortcuts of adiabaticity (STA) was proposed, to implement fast adiabatic-like control protocols \cite{erik2013,rev}. Since then, these have become powerful methods with applications in quantum information processing, and more generally quantum computing during the past decade. These methods include counter-diabatic (CD) driving~\cite{mustafa2003,mustafa2005}, also known as transitionless quantum driving \cite{Berry2009,Adolfo}, fast-forward approach \cite{shumpei2008,shumpei2010} and invariant-based inverse engineering \cite{xi2010,xi2011}. Particularly, local CD driving and its variations have been utilized for many-body spin systems, e.g., to find the ground state  \cite{kazutaka2017,sels2017,pieter2019,hartmann2019,passare2020,hatomura2021}.  Recently, the digitized counterdiabaticity for quantum computing has been proposed \cite{naren2021}  in the context of quantum gates and circuits. It is found that the inclusion of CD interactions shallows the circuit depth significantly, as compared to previous digitized adiabatic computing~\cite{Barends2016}. The advantage of this method has been analyzed in quantum annealing \cite{passarellipra}, quantum approximate optimization algorithm (QAOA)~\cite{parnav2021,jonathan2022,ychai2022,pranav2022}, and variational quantum eigensolver (VQE) \cite{ZZhan2021}, two typical quantum-classical hybrid algorithms, intending to find the ground state of a target Hamiltonian. 
Apart from this, the implementation of these methods has been also reported in various scenarios, such as factorization~\cite{naren2021factor} and portfolio optimization~\cite{NN.Hegade2021}. Moreover, a wide range of optimization problems has been further tackled using the drastic advantage, illuminating that CD driving severs as a non-stoquastic catalyst in the acceleration of quantum adiabatic algorithms \cite{naren2022}. 

With the advantages of the counterdiabaticity, a challenging task is to find the efficient CD terms, see Refs. \cite{steve2015,jhyao2021,ieva2022}. Instead of finding CD terms using spectral knowledge, adiabatic gauge potentials \cite{Kolodrubetz2017} have been proposed to achieve the approximate CD terms recently, particularly for many-body spin systems~\cite{sels2017,pieter2019,hatomura2021}. The approximate CD terms can be easily found by using the variational principle~\cite{sels2017} and Floquet engineering~\cite{pieter2019}, while diagonalizing the Hamiltonian for many-body spin system is difficult or time-consumption task to obtain the exact CD term. Along with this, an approximate form of the CD term is pre-selected, and the coefficients are determined by minimizing the action or the nested commutator (NC). However, more  high-fidelity control of larger spin systems requires  higher-order  NC of CD terms, beyond the two-body interaction \cite{naren2021}, which implies the problematic implementation with more CNOT gates. Actually, the optimal (local) CD driving can be achieved by hybridizing the technique of STA through optimal control theory \cite{steve2015,ieva2022} and reinforcement learning \cite{jhyao2021}.  

\begin{figure}[]
	\centering
	\includegraphics[totalheight=2in]{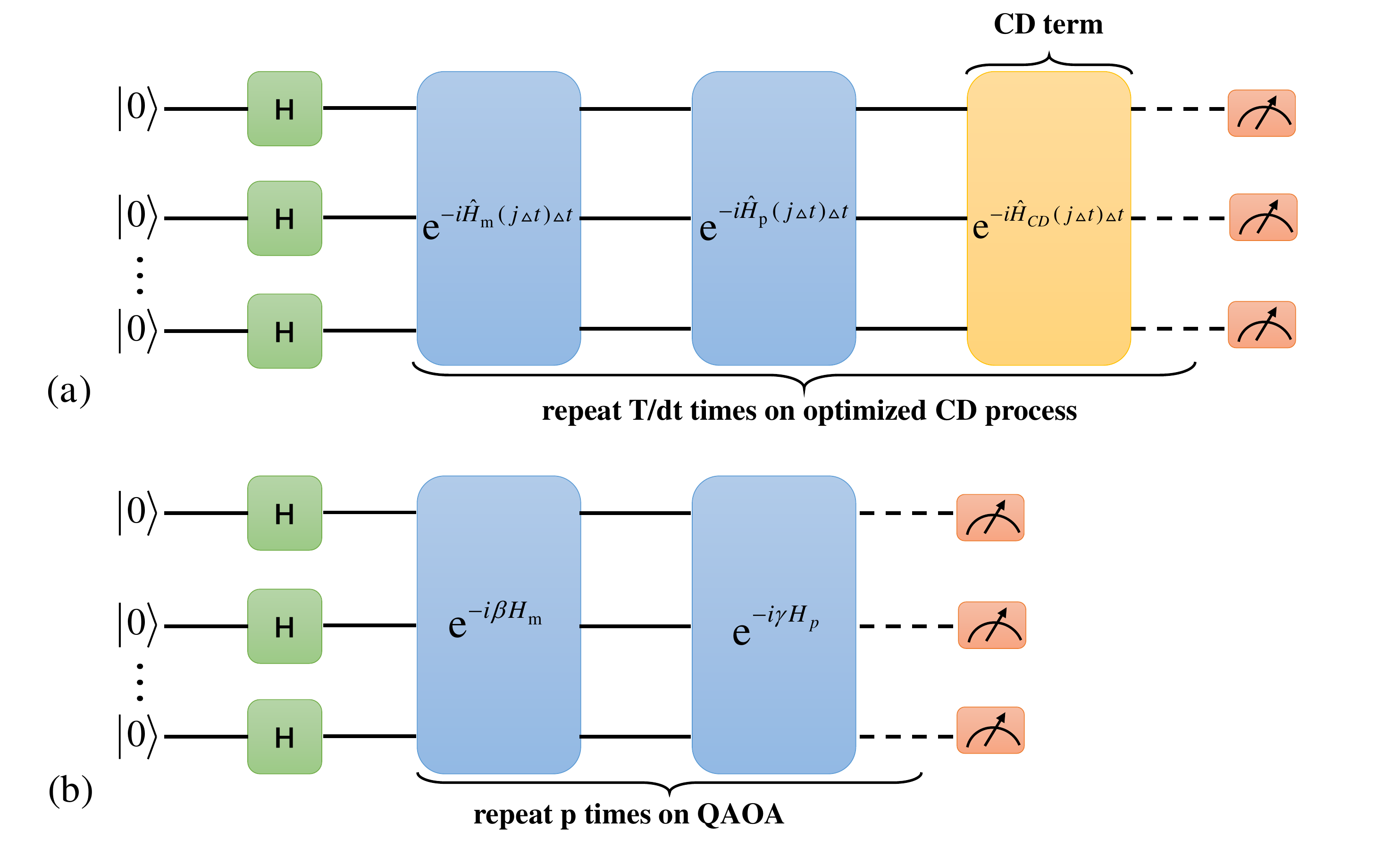}
	\caption{Schematic diagram for the digitized adiabatic algorithm assisted by CD term and (b) QAOA used here for the GHZ state preparation in the nearest-neighbor Ising model.} 
	\label{fig:circuit}
\end{figure}

In this work, we propose an alternative way to find efficient counter-diabaticity by using a variational quantum circuit, where quantum circuit and classical optimizer are applied to optimize parameters to minimize a cost function. In detail, we first replace the CD coefficients with free optimizable parameters and then use a classical optimizer to optimize them accordingly. The concrete example of
Greenberger-Horne-Zeilinger (GHZ) state preparation on the nearest-neighbor Ising model is further considered, see Fig. \ref{fig:circuit}, to demonstrate that the enhanced performance of optimal CD terms, as compared to digitized adiabatic evolution and QAOA.

The article is organized as follows. In the next Sec. \ref{sec2}, we explain the CD driving and classical optimizer in detail. In Sec.~\ref{sec3}, we investigate these methods with the nearest-neighbor Ising model and GHZ state preparation. Sec.~\ref{sec4} is devoted to the comparison to usual QAOA. Finally, the article ends up with brief conclusion in Sec. \ref{sec5}.

\section{CD driving and optimization}

\label{sec2}

The use of variational quantum circuit has bloomed a lot since most of them are hybrid quantum-classical algorithms that are considered one of the most crucial applications of the quantum computing paradigm in the NISQ era~\cite{John2018}. It consists of two parts, the quantum part that includes a circuit with variational parameters, and the classical part that includes iterative routines to optimize these parameters. Here, we will apply this method to counterdiabaticity, aiming to minimize a given cost function by tuning the free parameters of the quantum circuit to achieve optimal CD terms. 

In AQO, the trivial Hamiltonian $\hat{H}_m$ is prepared in its ground state and adiabatically evolved to the problem Hamiltonian $\hat{H}_p$. This can be encoded by the following Hamiltonian:
\begin{equation}
\hat{H}_0(t) = (1-\lambda(t) )\hat{H}_m + \lambda(t) \hat{H}_{p}, \label{con:Hamiltonian}
\end{equation}
where the time dependence of the system is introduced through the parameter $\lambda(t)$, and for the total time $T$, $\lambda(t)$ should meet the condition: $\lambda(0)=0$  and  $\lambda(T)=1$.
Typically, the adiabatic evolution takes a long time because of adiabatic criteria. For speeding this process up, we add a CD driving Hamiltonian $\hat{H}_{cd}(t)$, to suppress diabatic transitions, such that 
\begin{equation}
\hat{H}(t) = \hat{H}_0(t) + \hat{H}_{cd}(t), 
\label{con:Hamiltonian-2}
\end{equation}
where $\hat{H}_0(t)$ is the original Hamiltonian (\ref{con:Hamiltonian}) for the problem and $\hat{H}_{cd}(t)$ is the additional CD Hamiltonian. To elaborate the approximate CD term \cite{pieter2019,sels2017},
\begin{equation}
\hat{H}_{cd}(t)=\dot{\lambda}(t)\hat{A}^{(\ell)}_{\lambda}. \label{hcd}
\end{equation}
from the variational calculation, we write down the appropriate adiabatic gauge potential $\hat{A}_\lambda^*$ in terms of NC method \cite{pieter2019}, yielding
\begin{equation}
\hat{A}_\lambda^{(\ell)}=i\sum_{k=1}^{\ell}\alpha_k(t)\underset{2k-1}{\underbrace{[\hat{H}_0,[\hat{H}_0,\dots,[\hat{H}_0}},\partial_{\lambda}\hat{H}_0]]].  \label{con:A_lambda}
\end{equation}
Here the approximate gauge potential defined is determined by a set of coefficients, like $\alpha_1, \alpha_2, \cdots, \alpha_{\ell}$, and where $\ell$ determines the order of the expansion. 
The approximate CD coefficient $\alpha_{\ell}(t)$ can be finally obtained by minimizing the action $S_{\ell}=Tr[\hat{G}_{\ell}^2]$,
	with  the operator $\hat{G}_{\ell}$ being defined by $\hat{G}_{\ell}=\partial_{\lambda}\hat{H}_0-i[\hat{H}_0,\hat{A}_{\lambda}^{(\ell)}]$.

Based on the approximate CD term (NC=1), we use a variational circuit to find optimal CD parameters by the quantum-classical hybrid algorithm.  The circuit can be firstly constructed with gates to digitize the time evolution of Hamiltonian reads
\begin{equation}
U(0,T)=\prod_{j=1}^{n}e^{-i\hat{H}_0(j\Delta t)\Delta t} e^{-i\hat{H}_{cd}(j\Delta t)\Delta t},
\label{U_t}
\end{equation}
where $n=T/\Delta t$ is trotter step, $T$ is total evolution time and $\Delta t$ is the interval of every trotter step.
Here, $\hat{H}_{cd}(t)$ can be considered a non-stoquastic catalyst \cite{naren2022}. The unitary $U(0,T)$ generates an output state $\ket{\psi_{out}}$
	\begin{equation}
	\ket{\psi_{out}} = U(0,T)\ket{\psi_0},
	\label{U_t-0}
	\end{equation}
where $\ket{\psi_0}$ is the $N$ qubit ground state $\ket{+}^{\otimes N}$. 

On the other hand, for the classical optimization routine in quantum-classical hybrid algorithms, a variety of optimization algorithms have been proposed but a gradient-based systematic optimization of parameters is crucial. For that purpose, we use simultaneous perturbation stochastic approximation (SPSA), which is an algorithmic method for optimizing systems with multiple unknown parameters. It is appropriately suited to large-scale population models, adaptive modeling, and simulation optimization. The applications of SPSA are widely implemented in many problems, such as neural network training\cite{GCau1994}, statistical model parameter estimation and fault exposure\cite{alessandri1995}, adaptive control of dynamic systems\cite{spall1994,spall1997}, and using QAOA with the SPSA algorithm to solve the Max-cut problem\cite{salonik2021}. 

In general, there exist many classical optimizers, such as adaptive moment estimation algorithm (ADAM), constrained optimization by linear approximations (COBYLA), with their pros and cons. In contrast to others, SPSA uses only the objective function measurements of the objective function. So SPSA is considered to be efficient in high-dimensional problems in terms of providing a good solution for a relatively small number of measurements. More specifically, a differentiable cost function $C(\theta)$ in SPSA algorithm is considered, and $\theta$ is a L-dimensional vector. The optimization problem can be translated into finding a new $\theta$ at which $\frac{\partial C(\theta)}{\partial \theta}=0$. SPSA starts with an initial parameter vector $\hat{\theta}_0$, yielding
	\begin{equation}
	\hat{\theta}_{k+1}=\hat{\theta}_k-a_k\hat{g}(\hat{\theta}_k),\label{ak}
	\end{equation}
	where $\hat{g}_k$ is the estimate of the gradient $g(\theta)=\frac{\partial C(\theta)}{\partial \theta}$ at the iterate $\hat{\theta}_k$ based on prior measurements of the cost function, and $a_k$ is the size of step, a positive number. It is robust to any noise that may occur when measuring the function $C(\theta)$, therefore the function $y(\theta)=C(\theta)+\varepsilon$, where $\varepsilon$ denotes perturbation on the output. Then we can get the estimated gradient at each iteration step
	\begin{equation}
	\hat{g}_{ki}(\hat{\theta}_k)=\frac{y(\hat{\theta}_k+c_k\Delta_k)-y(\hat{\theta}_k-c_k\Delta_k)}{2c_k\Delta_{ki}},\label{ck}
	\end{equation}
	where $c_k$ is a positive number and each parameter is simultaneously perturbed by either $\pm c_k$. $\Delta_k=(\Delta_{k1},\Delta_{k2},\dots,\Delta_{kL})^T$ is a perturbation vector, which is generated by using a zero-mean distribution. 
In the above equations, $a_k$ and $c_k$ reduce over the optimization iterations to converge to a final result, defined as
	\begin{equation}
	a_k=\frac{a}{(k+1+A)^{0.602}}, \qquad   c_k=\frac{c}{(k+1)^{0.101}},\label{ac}
	\end{equation}
where $A$ is the number of iterations multiply 0.01, and $a$, $c$ are the initial values. Besides, $N_k$ is used to denote the number of iterations.
In what follows that we shall combine CD driving with the classical optimizer in the Ising model for GHZ state preparation, and compare the results with CD terms obtained with NC ($\ell=1$) and usual QAOA as well.

\section{Nearest-neighbor Ising model and GHZ state preparation}
\label{sec}
As a heuristic study, we consider the spin 1/2 Ising chain with nearest-neighbor interaction, which is of importance for many-body systems and quantum optimization problems~\cite{reviewNISQ, naren2021}.  The Hamiltonian of nearest-neighbour Ising model can be written as
\begin{equation}
\mathcal{H} = \sum_{i=1}^{N}h_z\sigma_z^i+\sum_{i=1}^{N}J\sigma_z^i\sigma_z^{i+1},
\label{ising}
\end{equation}
where $N$ is the system size, $J$ is coupling strength between qubits and $\sigma_z$ is the Pauli spin 1/2 operator, and the periodic boundary condition $\sigma^{N+1}=\sigma^1$ is assumed. In the ferromagnetic case $(J < 0)$ neighboring vertices prefer to align to the same spin, whereas the opposite happens in the antiferromagnetic setting $(J > 0)$. If $J$ is a random number and can either be positive or negative, then we have what is called a spin glass. The Ising antiferromagnet without an external field resembles a graph maximum cut (Max-Cut) problem that can be solved via combinatorial optimization, see Ref.~\cite{coja2022}.
Here we focus on the preparation of the entangled GHZ state, $\left|\mbox{GHZ} \right\rangle =(\left|0\right\rangle^{\otimes N}+\left|1\right\rangle ^{\otimes N})/\sqrt{2}$, as the  ground state of the Hamiltonian (\ref{ising}) in the ferromagnetic case of $h_z=0$ and $J=-1$. Our paradigmatic model is different from the transverse field Ising model, in which there exist quantum phase transitions and the finite-rate adiabatic passage across a quantum critical point can be accelerated by using CD driving by mean-field theory~\cite{adolfo2012}.

To encode the GHZ state preparation into the adiabatic quantum computing see Eq. (\ref{con:Hamiltonian}),
we write  down the time-dependent Hamiltonian,
\begin{equation}
\hat{H}_0(t) = (1-\lambda(t))\sum_{i=1}^{N}h_0\sigma_x^i+\lambda(t)\sum_{i=1}^{N}J\sigma_z^i\sigma_z^{i+1}, 
\label{eq:H0}
\end{equation}
where the function $\lambda(t)$ is, without loss of generality,  is assumed as
\begin{equation}
\lambda(t)=\sin^2
\left[\frac{\pi}{2}\sin^2(\frac{\pi t}{2T})\right].
\end{equation}
With the help of of NC method \eqref{con:A_lambda}, we  can calculate the  approximate CD  term. By keeping the first order ($\ell$ = 1), we have 
\begin{equation}
\hat{A}_{\lambda}^{(1)}=2\alpha_1(t)Jh_0\sum_{i=1}^{N}(\sigma_z^i\sigma_y^{i+1}+\sigma_y^i\sigma_z^{i+1}),
\end{equation}
from which we have  the CD terms, implying the two-body spin interactions, consisting of $\sigma_y\sigma_z$ and $\sigma_z\sigma_y$ terms. From ~\eqref{hcd}, the Hamiltonian of CD term can be further obtained as
\begin{equation}
\label{CD-Ising}
\hat{H}_{cd}(t)=\theta_{cd}\sum_{i=1}^{N}\left(\sigma_z^i\sigma_y^{i+1}+\sigma_y^i\sigma_z^{i+1}\right),
\end{equation}
with $\theta_{cd}=2\dot{\lambda}(t)\alpha_1(t)Jh_0$.
By minimizing the action $S_1=Tr[\hat{G}_1^2]$, where $\hat{G}_1=\partial_{\lambda}\hat{H}_0-i[\hat{H}_0,\hat{A}_{\lambda}^{(1)}]$, we finally achieve 
\begin{equation}
\alpha_1(t)=-1/16[(-1+\lambda)^2h_0^2+J^2\lambda^2].
\label{coeffs}
\end{equation}
It is emphasized again that higher-order NC suggests the higher fidelity for GHZ state preparation for the larger spin systems \cite{naren2021}. However, high-order NC implies the many-body interactions, beyond typical two-body interactions, which makes the physical implementation difficult, for instance, in photonic quantum circuits \cite{alberto2014}. Hence it is quite natural for us to focus on the optimization of CD terms with two-body interaction.

As a consequence, the output state is given by  $\ket{\psi_{out}}=U(\vec{\theta})\ket{\psi_0}$,
where 
the unitary operator $U(\hat{\theta})$ ~(\ref{U_t}) is
\begin{equation}
U(\vec{\theta})=\prod_{j=1}^{n}\prod_{k=1}^{N} e^{-i\theta_x(j\Delta t)\sigma_x^k\Delta t}
	e^{-i\theta_{zz}(j\Delta t)\sigma_z^k\sigma_z^{k+1}\Delta t}
	e^{-i\theta_{cd}(j\Delta t)(\sigma_z^k\sigma_y^{k+1}+\sigma_y^k\sigma_z^{k+1})\Delta t},
\end{equation}
where $\vec{\theta}=[\theta_x, \theta_{zz}, \theta_{cd}]$, with $\theta_x(\Delta t)=(1-\lambda(\Delta t))h_0$, $\theta_{zz}(\Delta t)=\lambda(\Delta t)J$ and $\theta_{cd}(\Delta t)$.
\begin{figure}[]
	\centering
	\begin{minipage}[t]{0.48\textwidth}
		\centering
		(a)	\includegraphics[totalheight=2in]{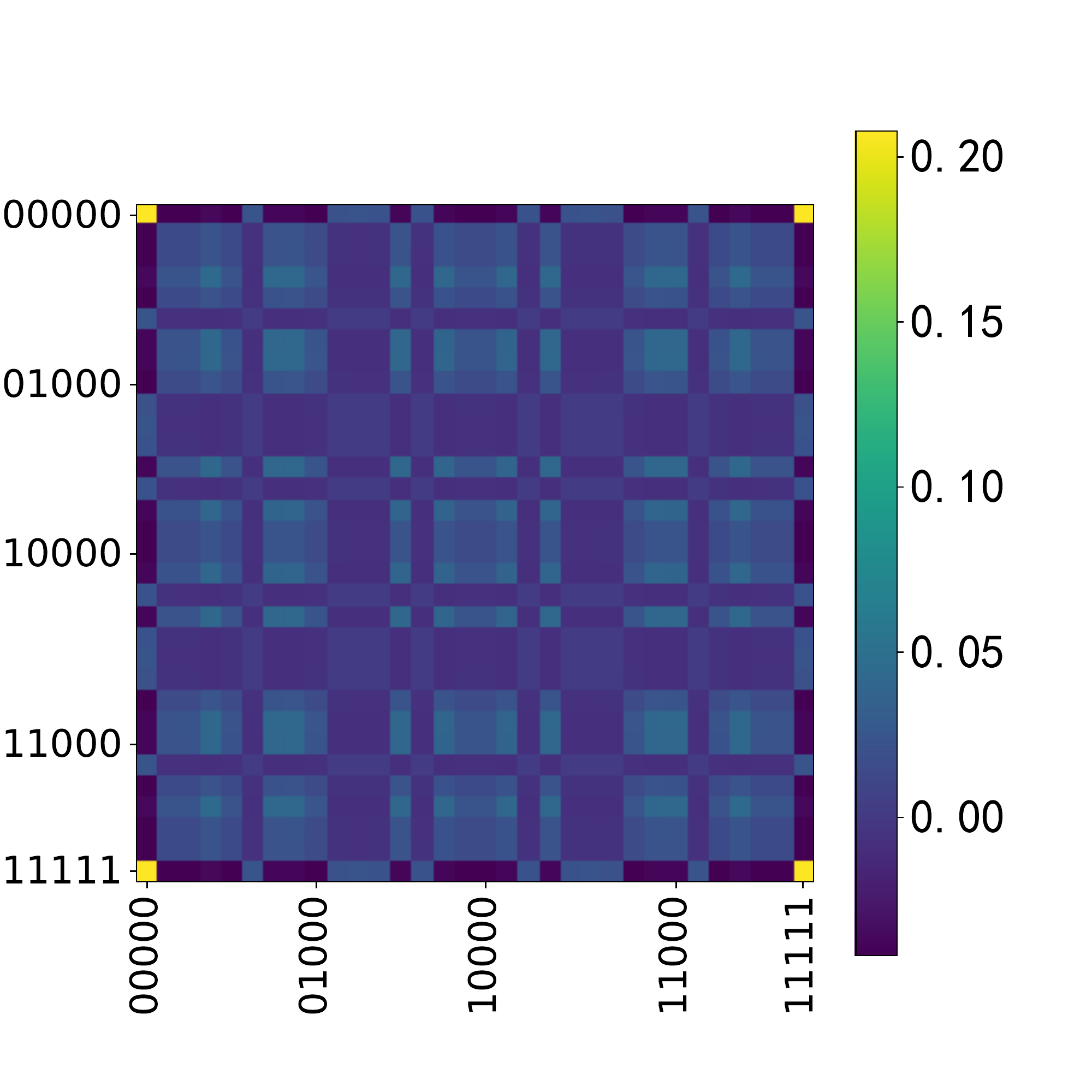}
	\end{minipage}
	\begin{minipage}[t]{0.48\textwidth}
		\centering
		(b)\includegraphics[totalheight=2in]{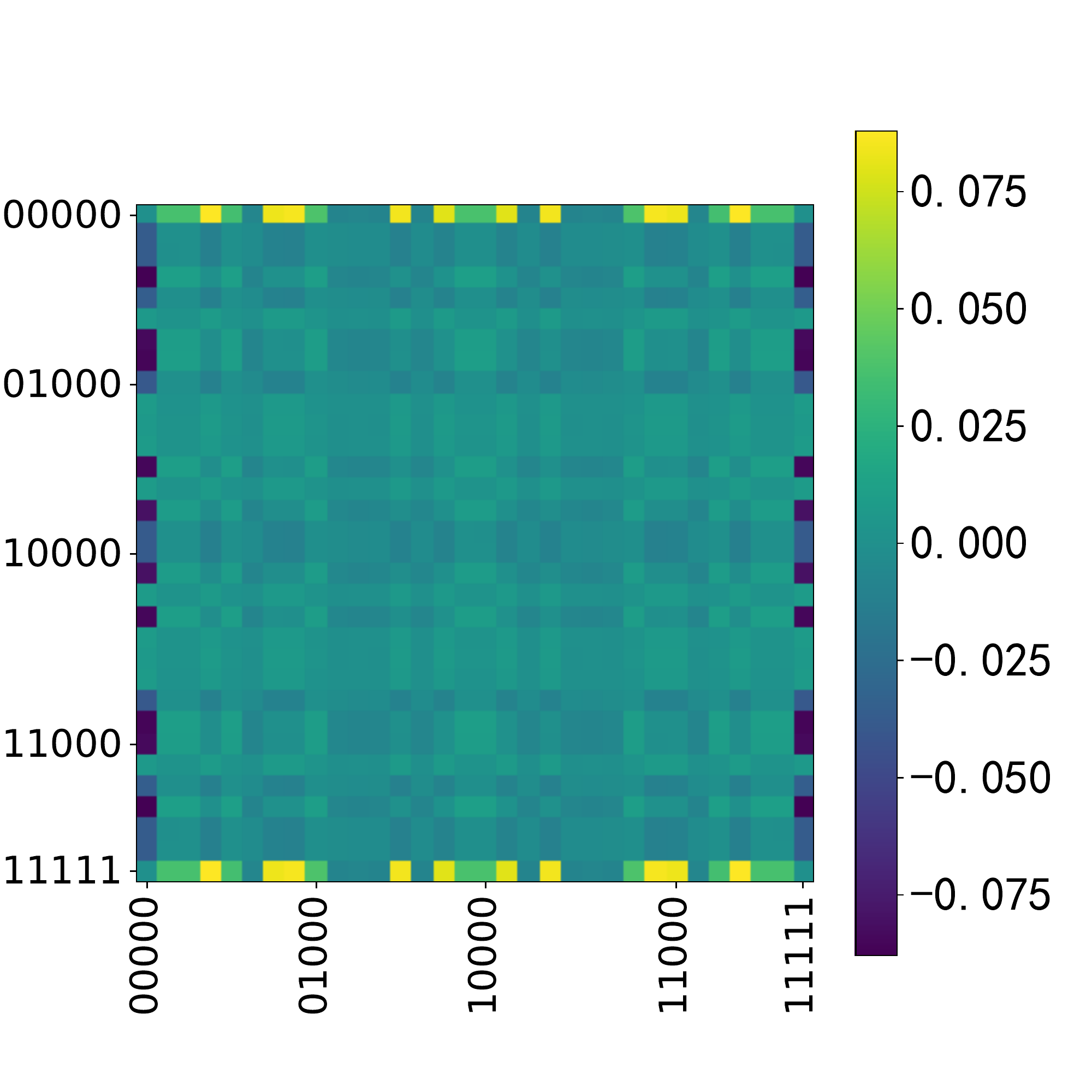}
	\end{minipage}
	\begin{minipage}[t]{0.48\textwidth}
		(c)	\includegraphics[totalheight=2in]{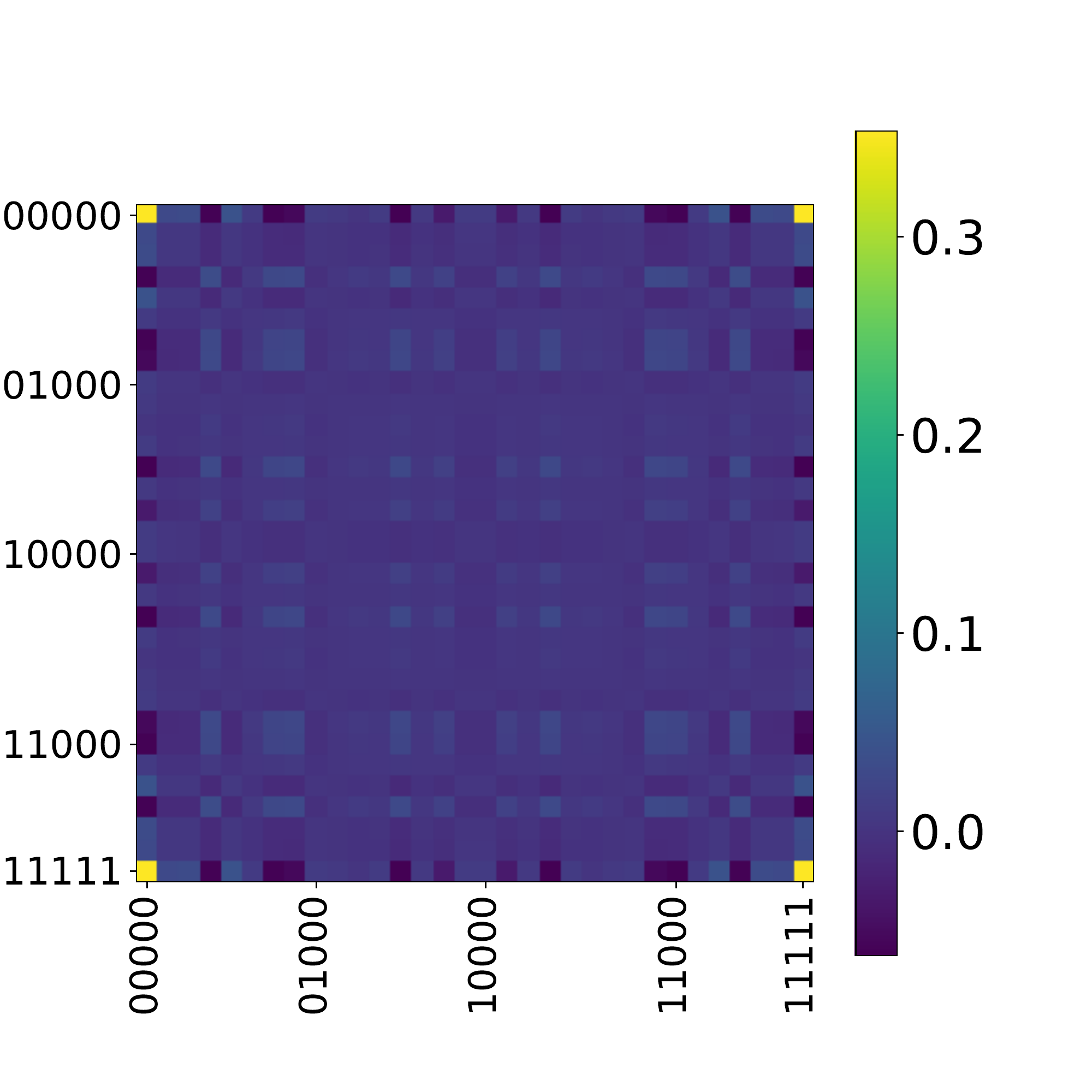}
		\centering 
	\end{minipage}
	\begin{minipage}[t]{0.48\textwidth}
		\centering
		(d)	\includegraphics[totalheight=2in]{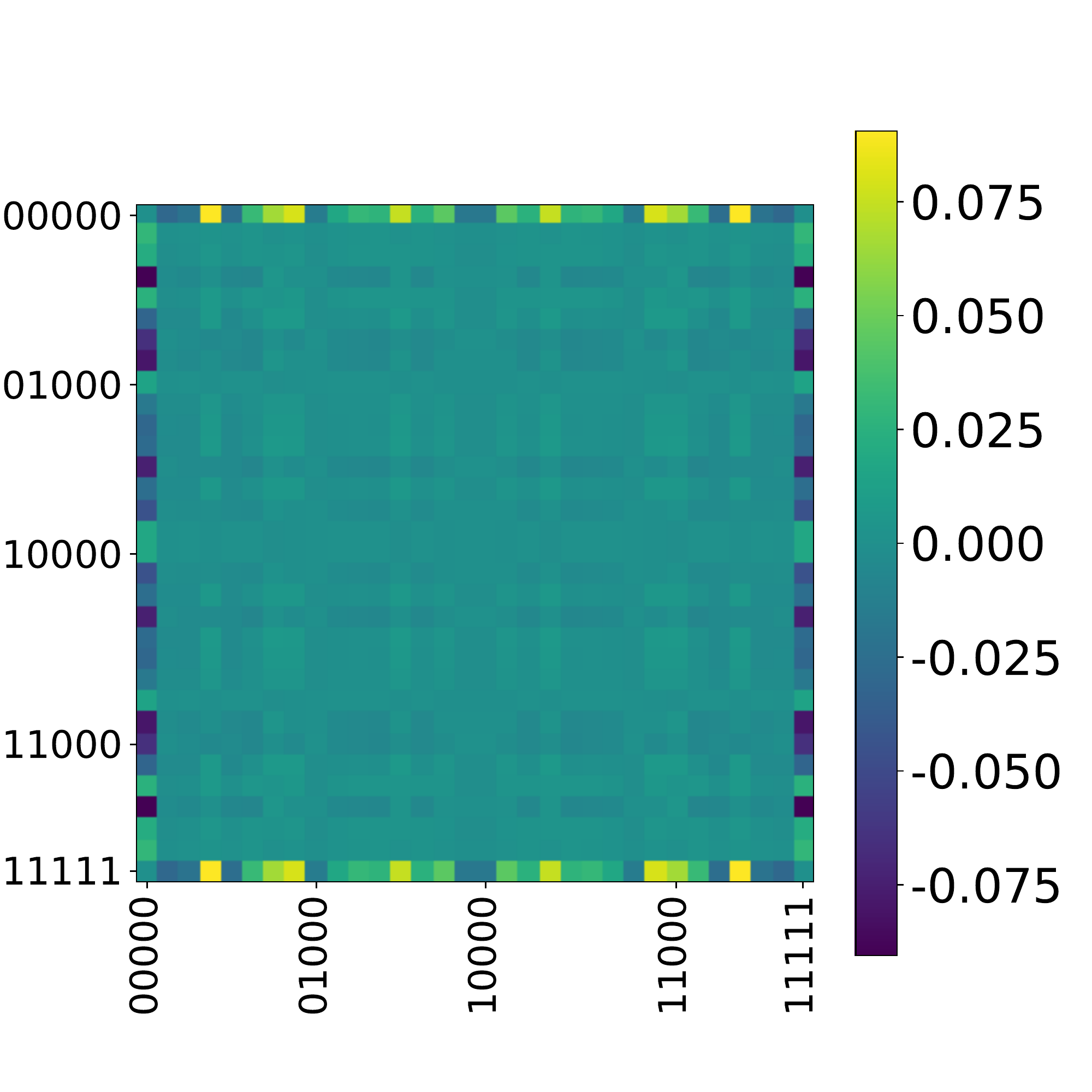}
	\end{minipage}
	\caption{Density matrix representation for 5-qubit GHZ state preparation (at $t=T$) using the approximate ((a,b))
		and optimal ((c,d)) CD terms, where the approximate CD terms are obtained NC ($\ell=1$), and CD is accordingly optimized by SPSA  with the parameters $J=-1$, $dt =0.2$, $T=1$ and $N_k=100$. (a,c) and (b,d) show real and imaginary parts, respectively. These results for 5-qubit system are calculated on an ideal quantum simulator.}  
	\label{fig:state0}
\end{figure}
In principle, one can digitize the CD term (\ref{CD-Ising}) with the coefficient (\ref{coeffs}) in the gate-based circuit to accelerate the adiabatic preparation of GHZ state. Instead,  we further propose a variational quantum circuit to optimize the digitized CD term for the same proposal, by considering  $\theta_{cd} (j \Delta t)$ as  free parameters. 
To this end,  the cost function, $C=1-\left |\left \langle\psi_{out}|\psi_{tar}\right\rangle\right|^2$,  is minimized for the target state $|\psi_{tar}\rangle=|\mbox{GHZ}\rangle$. As we discussed before, the SPSA algorithm is adopted to optimize the variational parameters $\theta_{cd} (j \Delta t)$, with the boundary conditions $\theta_{cd}(0)=0$ and $ \theta_{cd}(T)=0.$ Thus, the  previous CD term  (\ref{CD-Ising})  becomes
	\begin{equation}
	\label{cd-express}
	\tilde{\hat{H}}_{cd}=\tilde{\theta}_{cd}\sum_{i=1}^{N}\left(\sigma_y^i\sigma_z^{i+1}+\sigma_z^i\sigma_y^{i+1}\right),
	\end{equation}
to achieve higher fidelity, where the coefficient $\tilde{\theta}_{cd}$ is inferred from the optimal (site-independent) parameters $\tilde{\theta}_{cd} (j \Delta t)$, in the digitized quantum circuit, based on SPSA algorithm. 

In Fig.~\ref{fig:circuit} (a), we show the form of the quantum circuit,  where the parameters are either $\theta_{cd}$ or $\tilde{\theta}_{cd}$ optimized using SPSA. We will implement the results on an ideal quantum simulator based on Qiskit~\cite{qiskit}, where the number of iterations $N_k=100$,  $a=0.15$,  and $c=0.1$ are set.

\begin{figure}[]
	\centering
	\includegraphics[totalheight=2.5in]{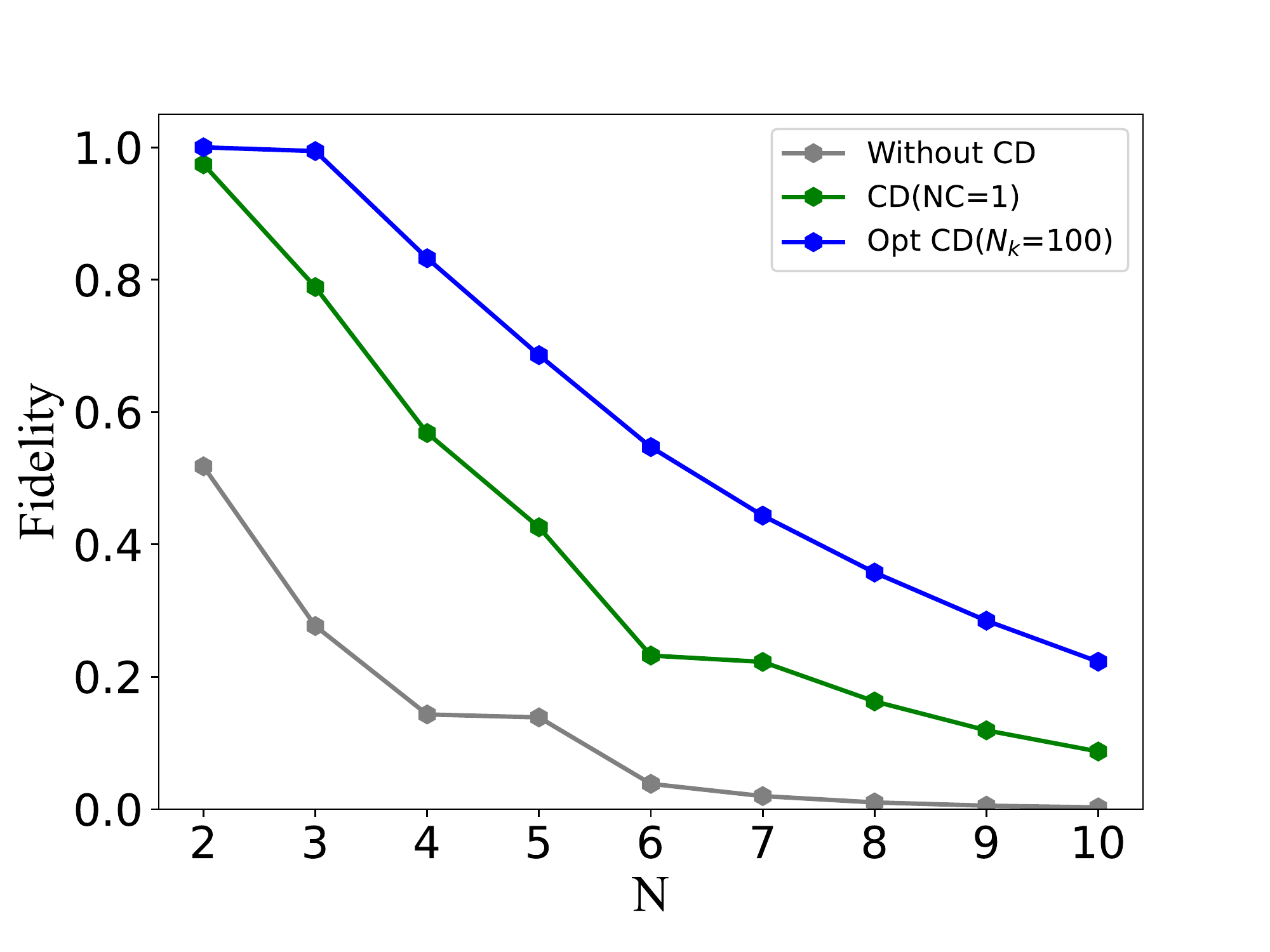}
	\caption{Fidelity of the prepared GHZ state in Ising spin systems with various sizes ranging from 2 to 10 qubits, where the approximate CD  NC ($\ell=1$) and its optimization by SPSA are included, and for completeness, the case without CD terms is also compared. The parameters: $J=-1$, $\Delta t = 0.2$, $T= 1$, and $N_{shots} = 1000$ implemented on ideal quantum simulator Qiskit.}
	\label{fig:2-10}
\end{figure}

Figure~\ref{fig:state0} displays the density matrix representation at $t=T$ for the GHZ state preparation in 5-qubit system with the approximate and optimal CD terms based on NC ($\ell=1$).
Obviously, it is exemplified in Fig.~\ref{fig:state0} that the results with both types of CD terms adapt to the target GHZ state for $J=-1$. However, Fig.~\ref{fig:2-10} further shows the optimized CD terms by SPSA have the better performance,
where the results with approximate and optimal CD terms based on NC ($\ell=1$) are compared with the normal adiabatic evolution. 
In detail, without CD driving, the final fidelity of the prepared GHZ states in a 5-qubit system is  quite low, due to the adiabatic error when total time $T=1$ is very short, and does not fulfill the adiabatic criteria. However, the approximate CD terms are not valid at all for the all cases of large spin size $N$ and interaction systems $J$~\cite{naren2021}. With the assist of SPSA optimizer, the optimal CD terms improve the fidelity.
So our method provides a simple but efficient approach to improving the fidelity of entangled GHZ state preparation by keeping the structure of two-body interaction. Of course, one can also extended  to higher-order NCs, with the many-body interaction involved.
Also,  other methods including genetic algorithms \cite{PRhegde2022} and reinforcement learning \cite{jhyao2021} is interesting to incorporate for further exploration with richer structures of CD terms. 

\begin{figure}[]
	\centering
	\includegraphics[totalheight=2.4in]{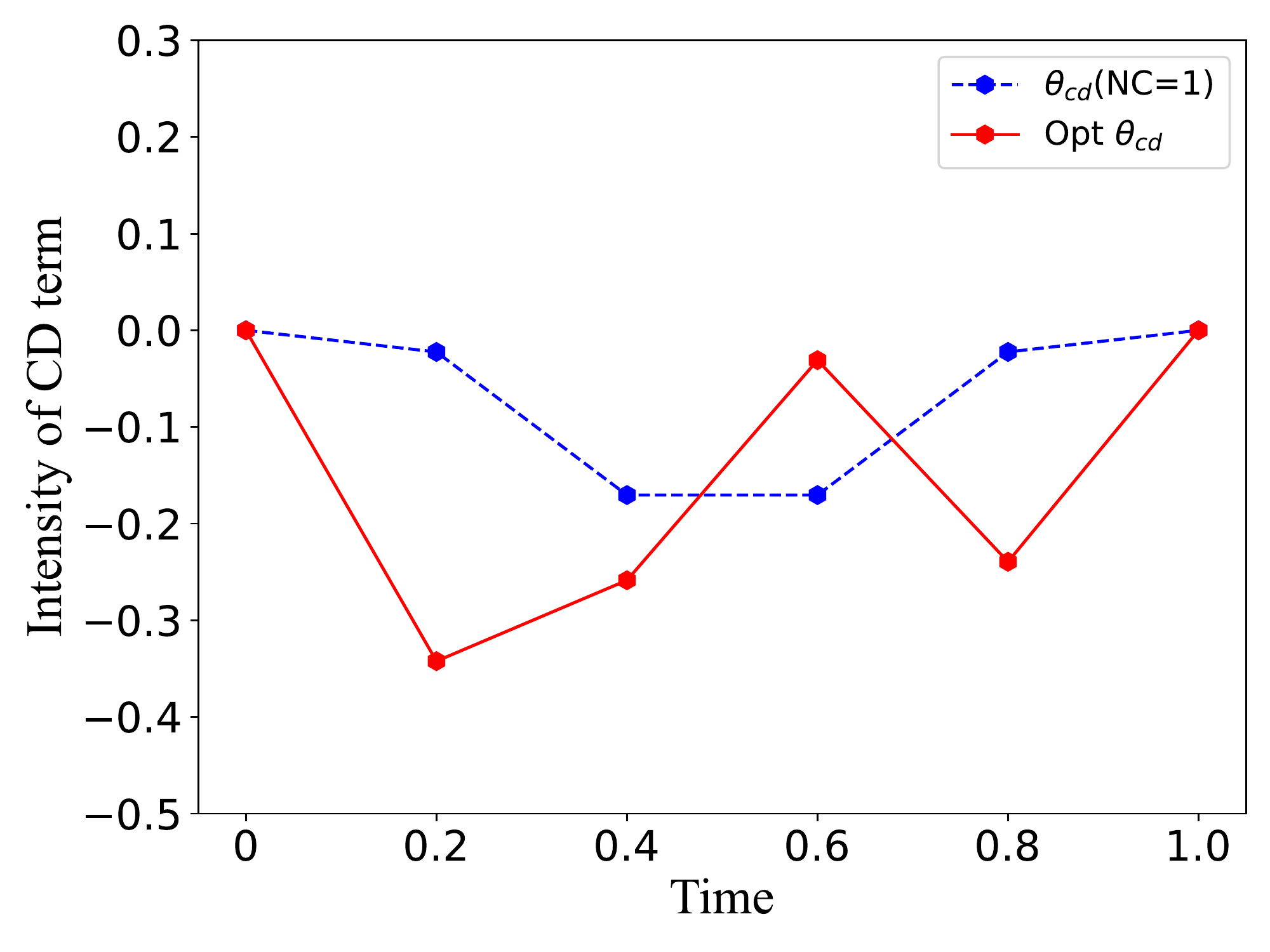}
	\caption{The coefficients of digitized CD terms for  the nearest-neighbor Ising model with 5 qubits, where  $\tilde{\theta}_{cd} (j \Delta t)$ (red line) is obtained by the optimal CD terms in Eq.~(\ref{cd-express}) and $\theta_{cd} (j \Delta t)$ (blue line) is obtained from Eq.~(\ref{coeffs}), the approximate CD term with NC ($l=1$).}
	\label{fig:cd-term} 
\end{figure}

\begin{figure}[]
	\centering
	\includegraphics[totalheight=2.5in]{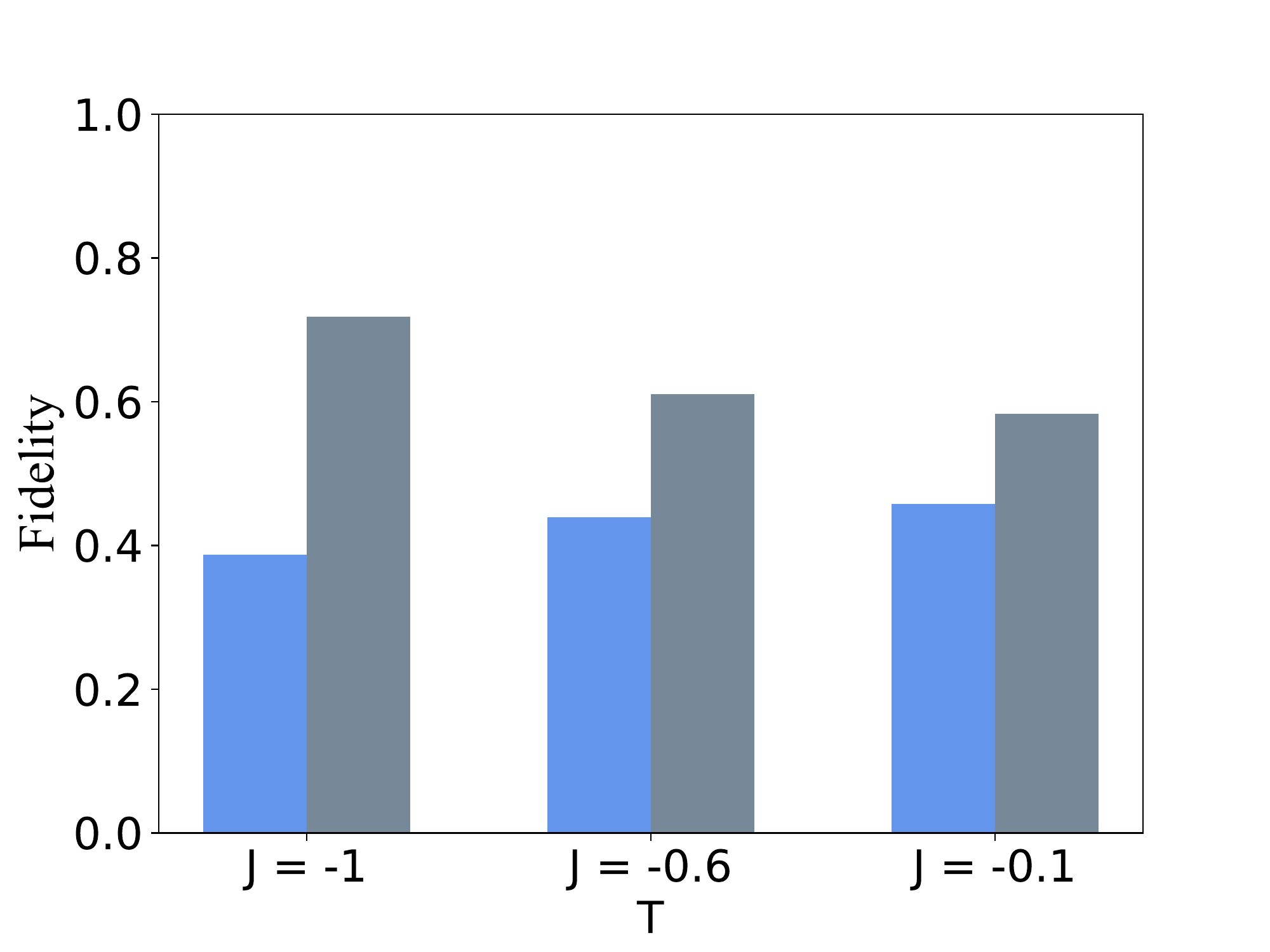}
	\caption{Fidelity of the Ising model with $h_z=0$ are compared for the approximate and optimal CD terms, where 
		blue and grey present the results from approximate (NC ($\ell=1$)), and corresponding optimized CD terms. $J=-1, -0.6, -0.1$ are considered in 5-qubit system, with $\Delta t=0.2$, $T = 1$, $h_0 = -1$, $N_{shots} = 1000$, $N_k=100$, $a=0.15$, $c=0.1$. Results were computed on ideal quantum simulator Qiskit.}
	\label{fig:Fidelity}
\end{figure}

Meanwhile, Fig.~\ref{fig:cd-term} illustrates the coefficients of digitized CD terms, $\theta_{cd} (j \Delta t)$ and $\tilde{\theta}_{cd} (j \Delta t)$, in the quantum circuit for the 5-qubit system, where their intensities are compared. As apparent, the fidelity is improved by the optimized CD terms, using SPSA algorithm, at the cost of the intensity of CD driving, proportional to the energy.  Finally, in Fig.~\ref{fig:Fidelity},  we check the fidelity of the entangle GHZ state preparation with various interaction $J$, where the rest parameters $J=-1, -0.6, -0.1$ and $h_z=0$ are considered in the Ising model with 5 qubits. It is concluded the results of  optimal CD terms outperform those of approximate one, with the same NC ($\ell=1$). 

\section{Comparison to QAOA}
\label{sec4}

Typically, QAOA is a quantum algorithm that attempts to solve such combinatorial problems~\cite{guerreschi2019,zhu2020,yang2016}. QAOA falls under the category of VQAs where $p$-layers of two unitaries $U_m(\beta)$ and $U_p(\gamma)$ are applied iteratively to an initial state $\ket{\psi_i} = \ket{+}^ {\otimes N}$, in the computational basis where $N$ is the system size. Thus the final state is given by
\begin{equation}
\ket{\psi(\gamma,\beta)} =  U_m{(\beta_p)}U_p{(\gamma_p)} \dots U_m{(\beta_1)}U_p{(\gamma_1)} \ket{\psi_i}, \label{eq:psi_beta_gamma}
\end{equation}
where $U_m(\beta)= e^{-i\beta \sum_{i=1}^N h_0\sigma^i_x}$ and $U_p(\gamma)=e^{-i\gamma \sum_{i=1}^NJ\sigma^i_z\sigma^{i+1}_z}$. $U_m(\beta)$ is known as the mixer unitary and $U_p(\gamma)$ is known as the problem unitary. ($\gamma$, $\beta$) are free optimizable parameters that are tuned by the classical optimizer to minimize the cost function $C^{'}$, that shows infidelity of the output state
	\begin{equation}
	C^{'}=1-\left |\left\langle\psi(\gamma,\beta)|\psi_{tar}\right\rangle\right|^2.
	\end{equation}

We come up with a perspective of QAOA from quantum control by having an eye on Eq.~(\ref{eq:psi_beta_gamma}). Indeed, QAOA represents the variational quantum control task that alternatively evolves the problem Hamiltonian $\hat{H}_p$ and the mixing Hamiltonian $\hat{H}_m$ for the operation time of $\gamma_i$ and $\beta_i$, respectively. Thus, it shares the same expression in Eq.~(\ref{eq:H0}) for our problem, where $\lambda(t)$ is no longer continuous, but a binary value controller $\lambda(t)=\{0, 1\}$ instead, reproducing a bang-bang control toward a target state. In this way, we reckon that it is necessary to compare our protocol, a digitized quantum annealing with counterdiabaticity, with QAOA by bounding the operation time, as a critical energetic quantity that affects the performances of bang-bang control and quantum annealing.

In the following, we compare the performance of QAOA ($p=1$ and $p=2$) with the optimized CD evolution in the respect of fidelity with a bounded time of the algorithms. We define the time $T'=\sum_{i=1}^{p}(\gamma_i+\beta_i)$ to compare the performance through preparing GHZ state for $h_0=-1$ and $J=-1$, where the target state $\ket{\psi_{tar}}=\ket{\mbox{GHZ}}=(\ket{0}^{\otimes N}+\ket{1}^{\otimes N})/\sqrt{2}$. Also, we apply the same SPSA  as a classical optimizer while performing QAOA, see Fig. \ref{fig:circuit} (b).

In Table.~\ref{QAOAtable}, we have compared the fidelities in the same time with the optimized CD driving, which shows the fidelities of QAOA are lower than the results from optimized CD terms at $T^{'}=T=1$. We have to emphasize that the shorter time is of significance to prevent the state evolution from decoherence, and gate errors. As expected, the fidelities decrease as the system size increases. Compared to hybrid quantum-classical algorithms like QAOA, the optimal CD works better with bounded time, where the parameters for SPSA are the same, $N_k=100$, $a=0.05$ and $c=0.05$. In fact, QAOA can  be also accelerated by adding the CD terms, $\alpha \hat{H}_{CD}$, as a richer ansatz in variational circuit \cite{parnav2021,jonathan2022}, which is beyond the scope of our paper. 

\begin{table}[h]
	\caption{Table on the fidelity comparison between two methods of optimal CD driving and QAOA at $T^{'}=T=1$.  }
	\centering\label{QAOAtable}
	\setlength{\tabcolsep}{1mm}
	\begin{tabular}{|c|c|c|c|}
		\hline
		Qubit Number& Optimal CD&QAOA (p=1)& QAOA (p=2)\\ \cline{1-4} 
		4 & 0.77 & 0.46  & 0.62  \\ 
		6 & 0.49 & 0.25  & 0.37   \\ 
		8 & 0.30 & 0.10  & 0.19  \\ 
		10 & 0.18 & 0.04 & 0.10  \\ \hline
	\end{tabular}
	\vspace*{-4pt}
\end{table}

Last but not least, we briefly analyze the errors in our protocols, resulting in the state preparation's infidelity. The adiabatic error, i.e., excitation induced by the non-adiabatic process, is not canceled perfectly, even if we have introduced optimized CD terms. It can be reduced by prolonging the operation time, or equivalently scaling up the mixing and problem Hamiltonian. The Trotter error is also introduced by the first-order Suzuki-Trotter decomposition \cite{TS} in our digital quantum computing paradigm. We realize that the Trotter error increases since either two of the mixing Hamiltonian, problem Hamiltonian, and the CD terms do not commute. However, the Trotter error is still in the scale of $O(\Delta t^2)$ by extending the Baker-Campbell-Hausdorff formula to the case of three components as
\begin{equation}
e^Ae^Be^C=A+B+C+\frac{1}{2}[A,B]+\frac{1}{2}[A+B+\frac{1}{2}[A,B],C]+\cdots,
\end{equation}
where $A=-i\hat{H}_m\Delta t$, $B=-i\hat{H}_p\Delta t$, and $C=-i\hat{H}_{cd}\Delta t$.  Actually, the detailed error analysis can be worked out separately using Trotter formulas  \cite{Changhao}, with/without CD terms involved. 

\section{Conclusion}

\label{sec5}

In summary, we have implemented a variational quantum circuit to find optimal coefficients of CD driving, in order to speed up digitized adiabatic quantum computing. By suppressing non-adiabatic transitions but keeping high-fidelity, we have exemplified this method for the GHZ entangled state preparation in the nearest-neighbor Ising model and within a short time. The results have demonstrated that our method is superior to approximate CD term, based on NC, by minimizing the action.  With the first order NC ($\ell=1$),  the coefficients of two-body interaction have been optimized by SPSA, a classical optimizer with variational quantum circuits. In addition, our method is still better than the usual QAOA in terms of fidelity with bounded time, with reminiscences of the comparison between shortcuts to adiabaticity and optimal control theory with respect to robustness, operation time and etc. \cite{Entropy}. Of course, by using a richer ansatz of CD terms (e.g. with N-body interaction) one can also further increase the fidelity in quantum circuits, but the physical implementation becomes tough with the state-of-the-art architecture of NISQ devices. Moreover, other techniques of optimization, such as reinforcement learning \cite{parnav2021}, greedy and genetic algorithms \cite{PRhegde2022} are worthy of further exploration and comparison. Finally, it would be interesting to apply this method for various complex systems \cite{reviewNISQ}, e.g. Ising antiferromagnet model and transverse-field Ising model, solving the combinatorial optimization problems, and addressing the fundamental and relevant issues on the circuit complexity, gate  error mitigation, Trotter and adiabatic errors. 
\\
\\
\section*{Acknowledgement}
We acknowledge the discussions from Yongcheng Ding and N.N. Hegade. This work is partially supported from NSFC (12075145), EU FET Open Grant Quromorphic (828826) and EPIQUS (899368),  QUANTEK project (KK-2021/00070),  the Basque Government through
Grant No. IT1470-22 and Ministerio de Ciencia e Innovaci\'on (PID2021-126273NB-I00). 
X.C. acknowledges the Ram\'on y Cajal program (RYC-2017-22482).

\bibliographystyle{apsrev}

\end{document}